# RESERVOIR CHARACTERIZATIONS BY DEEP-LEARNING MODEL: DETECTION OF TRUE SAND THICKNESS

Ping Lu, Yanyan Zhang, Hua Yu, Stan Morris
Anadarko Petroleum Corporation, Houston, TX, 77380

## ABSTRACT

It is an extremely challenging task to precisely identify the reservoir characteristics directly from seismic data due to its inherit nature. Here, we successfully design a deep-learning model integrated with synthetic wedge models to overcome the geophysical limitation while performing the interpretation of thickness of sand bodies from a low resolution of seismic data. Through understanding and learning the geophysical relationship between seismic responses and corresponding indicators for sand thickness, the deep-learning model could automatically detect the locations of top and base of sand bodies identified from seismic traces by precisely revealing the lithology distribution. Quantitative analysis and extensive validations from wedge models and field data prove the robustness of the proposed methodologies. The true sand thickness, identified from the deep-learning model, provides an extremely useful guidance in enhancing the interpretation of lithological and stratigraphic information from seismic data. In addition, the proposed deep-learning approach eliminates the risks of over- and under-estimation of net-to-gross with a significant improvement with respect to the accuracy.

## INTRODUCTION

The fact that seismic has low resolution features due to frequency decay in great depth makes it incapable to capture the true thickness of sand bodies. Therefore, lithology cannot be precisely inverted and inferred from seismic attributes. However, obtaining a direct indicator of true sand thickness is always significant in understanding reservoir characterization, and precise estimation of net-to-gross could help the optimization of well planning Meza et al. (2015). Ricker (1953), Widess (1973), and Kallweit and Wood (1982) performed extensive researches on understanding of seismic vertical resolutions. Top and bottom interfaces for thin beds cannot be accurately resolved due to the features that most of the reservoirs are relatively small in the vertical dimension, and many of them are below the seismic resolution ($<\lambda/4$). Some of the earliest research, performed by Meckel and Nath (1977), Neidell and Poggiagliolmi (1977), and Schramm et al. (1977), talked about how to determine bed thickness below tuning. In fact, below tuning, the time separation of the reflections from the top and bottom of a reservoir essentially does not change. Estimating its thickness from the separation of trough and peak seismic loops results in a significant overestimate. Thus, the thickness must be determined by other methods. Connolly (2007) elaborated an approach by employing a map-based compensation of tuning effects followed by direct calibration to the reservoir property. Simm (2009) indicated that net pay prediction using band-limited impedance is more accurate than reflectivity. It is also demonstrated that various techniques have their own limitations and require certain assumptions that may not be fully met for determining the thickness of sand bodies, even though the fundamental understanding of seismic amplitudes and time separation of reflections from the top and bottom of sand bodies have been well studied. Therefore, conventional approaches sometimes cannot be fully trusted and accurate from an interpreter without having fundamental knowledges and fruitful experiences.

In order to overcome these geophysical limitations during the determination of thickness of thin beds from seismic responses, a deep-learning model is proposed to determine the precise locations where impedances differ between rock layers for the subsurface, rather than deriving absolute value of impedance. The output from a deep-learning model, indicating a strong link between seismic responses and lithology, tends to provide an accurate estimation about the surface from top to base of the sand bodies. In this paper, we will discuss how to develop a deep-learning-model based approach to effectively eliminate the tuning and interference effects by revealing the true thickness of reservoir through several examples.

## GEOPHYSICAL LIMITATIONS

### Interpretation of True Thickness of Thin Beds

Traveltime between zero-crossings of the main lobe on a minus 90 degree phase-rotated seismic attribute could be properly used for the measurements of actual bed thickness if the sand is within the marginal thickness ($\lambda/4 - \lambda$). For a low-impedance sand reservoir embedded in high-impedance shales, the typical response from seismic signal to a thin bed primarily consists of a single trough event. The center of the bed aligns with the maximum negative amplitude (λ/4 - 3λ/4) or with the center of the composite trough waveform (3λ/4 – λ). However, for a thinner reservoir, specifically thinner than λ/4, due to the interference of the reflectivity from both the top and base of reservoir, reading from the time-lag between zero-crossings, which are usually defined as apparent thickness, does not reflect the true thickness. In this case, the criteria to determine the true thickness for a thinner reservoir could be relatively challenging. The readings, if merely taking consideration of a linear link from seismic zero-crossings to the top and bottom of the bed, are not precise anymore. In fact, the thickness observed from the seismic response is much thicker than the true thickness of the reservoir, and it usually ends up having an overestimated consequence.

A synthetic wedge model could be used for quantitative interpretations from seismic data to study the tuning thickness, vertical resolution, traveltime shift below tuning thickness, and amplitude decay below the tuning thickness (Simm, 2009; Roden et al., 2017). Here, a typical wedge model, consisting of two different materials, is generated in the Figure 1. The top and bottom of the wedge, with opposite reflection polarities, are composed of shale layers, and the sandstone is present in the middle part. A thickness ranging from seismically thin (<λ/4) to marginally thick (λ/4 – λ) until seismically thick (>λ) are thoroughly covered in the model Zeng et al. (2005).

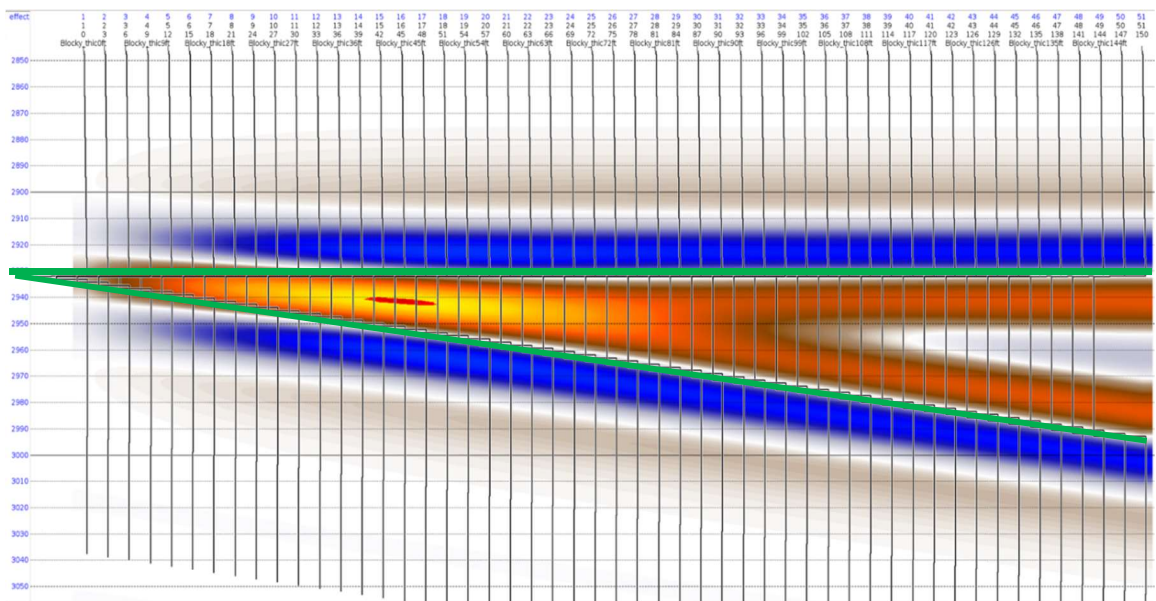

*Figure 1: Wedge model with indicators (marked in green color) reveals the top and base of sand thickness.*

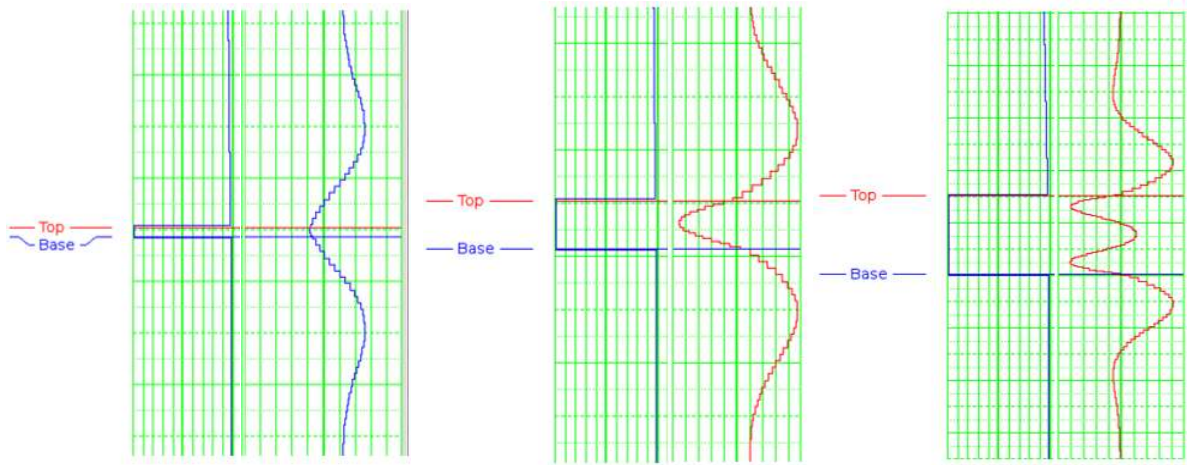

*Figure 2: Top and base of sand thickness corresponding to seismic amplitude.*

Three representative locations are selected to elaborate how to associate the top and base of sand with corresponding seismic reflectivity. Figure 2(a) presents the region where sand thickness is much less than a quarter of the seismic wavelength. Evidently, there is no distinct reflection indicating the true thickness. The apparent thickness reading from zero-crossing on seismic trace introduces an overestimated sand thickness. Figure 2(b) represents the marginal thickness region, where zero-crossings on seismic trace correspond to the boundaries of true sand thickness quite well. It turns out the apparent thickness reflects the true sand thickness very well. Figure 2(c) represents the region where sand is thicker than one seismic wavelength. Due to the interference where the end of the upper wavelet overlaps the top of the lower wavelet, it is quite challenging for an interpreter to distinguish two reflection coefficients given a trough doublet. The consequence is that the apparent thickness may be easily overestimated versus true sand thickness. These three typical locations have been chosen to train the deep-learning model in the following sections.

## METHODS

### Design of Deep-Learning Model

A deep-learning model is designed to provide us the opportunity to convert seismic trace directly to the lithological profile, which is highly desirable. Seismic response from a wedge model, with corresponding indicator representing thickness of sand bodies, form a pair as the input of the deep-learning model. Then, deep-learning model learns the physics of understanding how to link the sand thickness to the corresponding seismic responses based on selected pairs from the wedge model. Not surprisingly, the output of the deep-learning model consists of a series of accurate sand thickness, which perfectly coincide with the geologically defined sands automatically adjusted and inferred from the seismic trace by the trained deep-learning model. As a result, seismic polarity for the bed thinner than λ/4 and significant intervenes from seismic wavelets are not solely and directly used for determinations of sand thickness, instead inferences, learned from the combination of wedge model and deep-learning model with correct indicators of top and bottom of sands, are made for the entire wavelengths. It provides an effective solution to the challenging problems in terms of how to directly link seismic amplitudes to true sand thickness. In addition, although the deep-learning model is trained with synthetic wedge model with perfect condition, it has been successfully applied to infer true sand thickness on field data where significant noises, large erosions, and other interferences may exist.

## Conditional Generative Adversarial Networks

Motivated by the promising results of generative adversarial networks (GANs) in a variety of image processing and interpretation tasks Lu et al. (2018), we explore the potential of conditional GANs (cGANs) Isola et al. (2017) for overcoming limitation of seismology, as part of the overall goal of investigating the feasibility and performance of cGANs for detection of true sand thickness.

Every seismic trace is converted to individual spectrogram, and the indicator of sand thickness is also converted to an image as the inputs of the networks. The cGANs consists of two networks: a generator, that tries to generate a synthetic indicator based on the input seismic trace, and a discriminator, that tries to distinguish between synthetic indicators provided by the generator and real ones from the inputs by using the corresponding seismic trace as a condition. Both the generator and discriminator are trained in an adversarial manner. The performance of the cGANs is evaluated in terms of perceptual evaluation of images' quality between the mapping.

Specifically, let $x$ denote the conditioned input seismic spectrogram, $y$ be the desirable output of the network and $z$ be the Gaussian random noise, the objective of the purposed cGANs architecture is

$$\min_G \max_D \{E_{(x,y)}[\log D(x,y)] + E_{(x,z)}[\log(1 - D(x, G(x,z)))] + \lambda \cdot E_{(x,y,z)}[\|y - G(x,z)\|_1]\},$$

such that $D(\cdot)$ denotes the discriminating process, $G(\cdot)$ denotes the generating process and $\lambda$ represents the regularization coefficient. Here the regularization term controls the generated synthetic indicator to be similar with the original inputs, and $L_1$ norm is chosen to reduce the blurriness caused by squared distance. Figure 3 demonstrates the architecture of the cGANs used in this study.

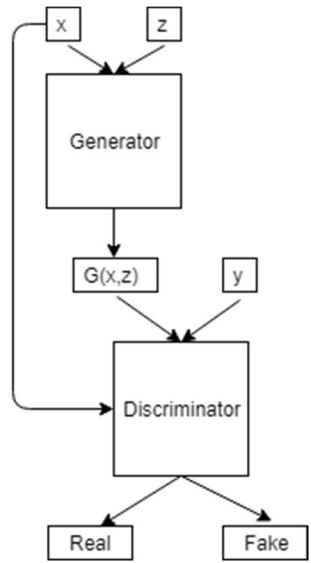

Figure 3: Graphical illustration for cGANs

The architectures used in this work are based on deep convolutional GANs that successfully tackles training instability issues when GANs are applied to high resolution inputs. Four key ideas are used to accomplish this goal. First, batch normalization is applied to most of the layers. Second, the networks are designed to have no pooling layers. Then, nonlinear ReLu is selected as activation function. Finally, the training is performed adopting the Adam optimizer.

The cGANs framework is designed to learn a physical mapping between seismic trace spectrograms and thickness indicators. To our knowledge, this is the first time as applying cGANs for detection of true sand thickness to overcome limitations of geophysics.

# RESULTS

## Validation on Synthetic Wedge Model

Efforts are first kicked off for applying the proposed neural network on the synthetic wedge model, as a successful recovery of wedge shape implies the capability to tune the seismic responses and detect the true sand thickness. In our experimentally synthetic model, there are 51 seismic traces with corresponding indicators representing the top and base of sand bodies, while the tuning thickness is at trace No. 15.

To make the training process representative, 4 out of the 51 pairs shown in Figure 4(a) are selected as the training data, where two pairs are selected from the region below tuning thickness, one pair from the marginal thick area, and the last pair from the seismically thick region. With these four pairs serving as the training inputs for cGANs, sand thickness is inferred for the rest of the seismic traces. Figure 4 shows the synthetic wedge model used for training and validation, and the predicted results for all the seismic traces from the wedge model are shown on the right. Despite some minor noises, the results show a clear wedge shape by correctly labeling the top and base locations of sand body regardless of the tuning effects and the interferences of overlapping seismic wavelets, which grants more confidences to proceed the experiment on more complex and challenging field data.

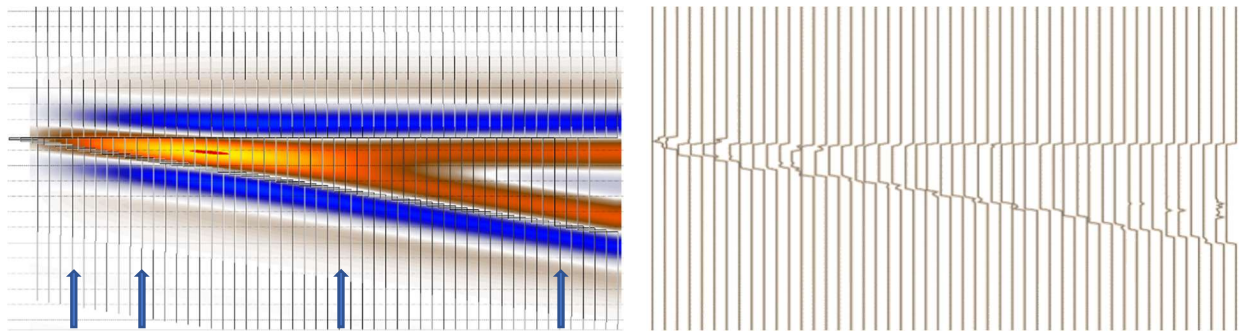

*Figure 4: a) Synthetic wedge model used for training and validation. b) Deep-learning inferred sand thickness for wedge model.*

## Inference on Field Offshore Seismic Data

Following the validation from synthetic data, a continuous work of applying the proposed deep-learning workflow on an offshore field data has been performed to further validate the robustness of the model. As the main objective of the effort is to detect true sand thickness, it is crucial to train the neural network with ground-truth information derived from the field data. Therefore, a particularly synthetic wedge model based on one typical well log is generated, from which four pairs of data, including the seismic traces as well as indicators with the top and base of sand, are utilized as the training inputs.

With the trained cGANs, the sand thickness for the entire field could be inferred instantly with millions of traces. Figure 5 presents two horizons (marked with blue and red color) indicating top and base of sand bodies interpreted from seismic responses completely based on zero-crossings for one reservoir layer. Inferences indicating thickness of sand bodies from deep-learning model are marked in yellow color, and the background region is identified as shale. It is apparent that inference from the deep-learning model does not always follow the zero-crossings on seismic trace where top and base of sand bodies has been correctly tuned according to the training information based on the wedge model. The comparisons, shown in Figure 5(b,c), indicate that sand bodies, inferred from cGANs, are more precise and detailed by comparing with interpreter's horizons.

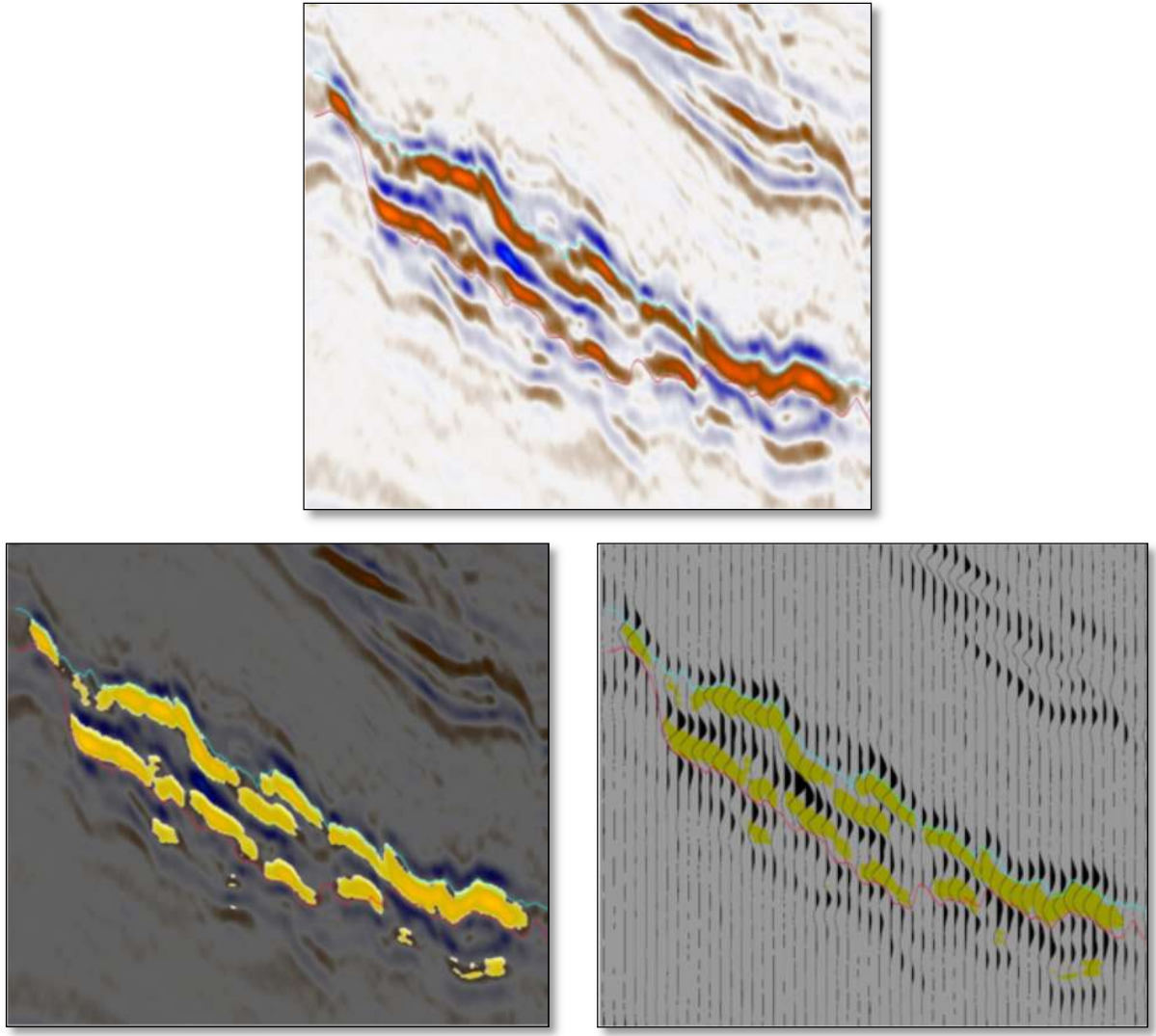

*Figure 5 a) Real seismic image with horizon information. b) Inferences from deep-learning model overlaid on seismic. The background region is considered as shale, where the yellow region is the inferred sand bodies. c) Inferences overlaid on seismic wiggles, which does not always follow the zero-crossing but instead, it tunes sand thickness based on its wedge model.*

It is noticeable that true sand thickness could be easily and heavily over- and under-estimated if zero-crossings are used for criteria of determinations. When it is in marginal thin region (shown in Figure 2(a)), apparent thickness indicated at seismic trace needs to be tuned thinner, whereas when it is in the seismically thick region (shown in Figure 2(c)), apparent thickness indicated at seismic trace needs to be tuned thicker. Figure 6 shows one region where the deep-learning model adjusts the seismic thickness to both scenarios by thickening and thinning of sand thickness identified from seismic trace, which demonstrates the solid evidences that the deep-learning model has learnt from training with the wedge model by gaining geophysical insights from seismic responses regarding corrections of below and above tuning thickness.

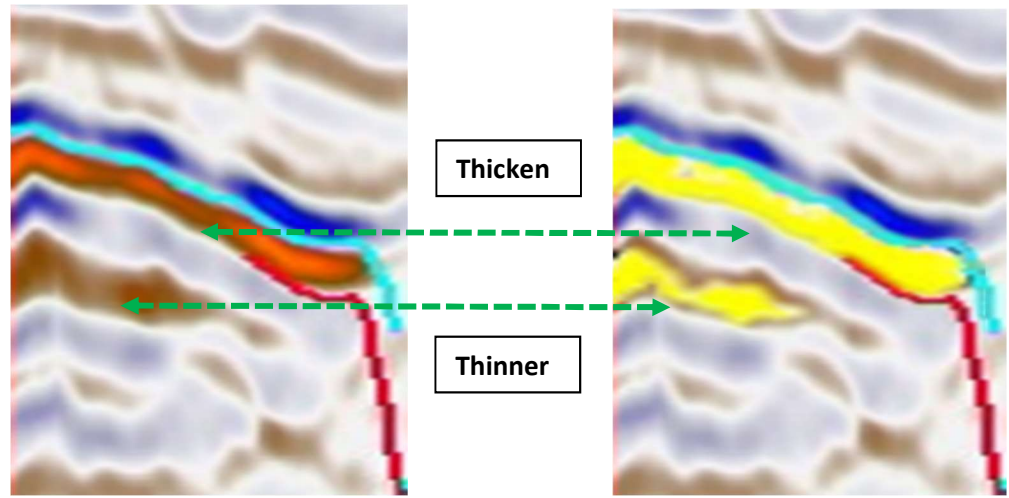

*Figure 6: Right image shows the inferred sand body (yellow region) corresponding to the left seismic, and it indicated that the neural network tunes seismic response in either direction to reflect the true thickness.*

Isochrones, which indicate changes of thickness in-between two horizons within a reservoir layer, are used as an alternative way to compare the sand thickness from the interpretations of zero-crossings of the original seismic versus the inferences from the deep-learning model. Due to the large scale and complexity of seismic volume, it takes tremendous efforts from geophysicists to precisely identify geological horizons, and to eventually generate isochrones for determining sand thickness correspondingly. More importantly, the accuracy of interpretation from seismic attributes could be heavily biased from each interpreter with different criteria and level of experiences, as well as affected by the tuning and interference effects, which fully mislead interpreter's interpretations and prevent the generated isochrones from being as an unambiguous estimation of reservoir thickness. However, the proposed deep-learning network could more effectively handle a large 3-D seismic volume with consistent criteria as well as to infer thickness by avoiding tuning effects in a totally automatic manner with much higher accuracy than human. Thus, the isochrones generated from deep-learning outputs provide finer details and higher resolutions, being able to draw a clear portrait for sediment deposition of true thickness as shown in Figure 7.

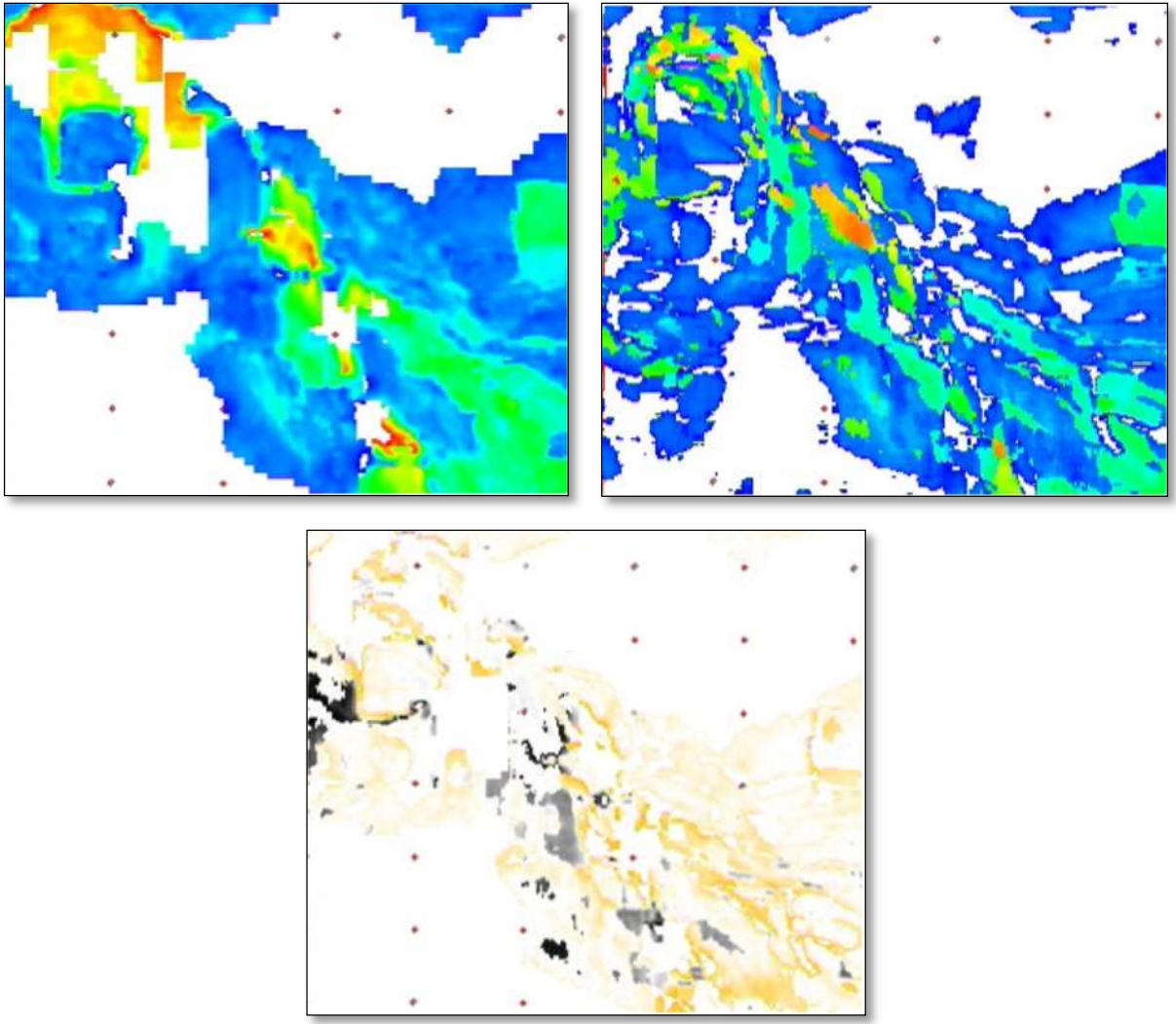

*Figure 7: a) Isochrones generated from an interpreter based on seismic information. b) Isochrones generated from deep-learning. c) Difference between two isochrones. Compared to interpreter's result, neural network generated the isochrones with more details and higher resolutions.*

Eventually, the proposed deep-learning approach would be entirely endorsed if it fully honors the log information. There are multiple well logs in the experimental region and all of them are used for validation purposes. For illustration purpose, one of the validation results is shown in the paper, and Figure 8(b) indicates the location where a well log is overlaid in the center of a seismic image. The corresponding seismic trace marked in blue color is shown in the Figure 8(a). It is confirmed that the apparent thickness reading from the seismic trace at log location does not reflect the true sand thickness ending up an underestimated consequence, if zero-crossings are used for deterministic criteria. However, the deep-learning model correctly infers the top and base of the sand, marked with black curve in Figure 8(a), and they match the log information quite well. Continuous sand geo-bodies corresponding to each seismic trace in Figure 8(b) are inferred from deep-learning model and shown in Figure 8(c).

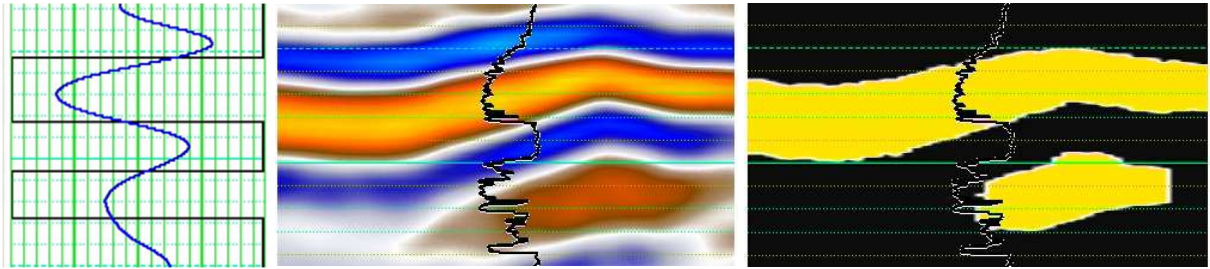

*Figure 8: Log information is used to valid the inference. a) Seismic trace on log location with inferred sand thickness. b) Seismic image with log data. c) Predicted binary (sand/no-sand) result with marked log location.*

## DISCUSSIONS

We have conducted some researches in investigating how many traces used for training the deep-learning model may eventually influence the performance of inferences. To avoid overfitting issue as well as improve the training efficiency, only small number of pairs consisting of seismic traces and corresponding sand thickness from wedge model are picked for training purpose. Then, some challenges of training such deep-learning model may be raised about how to select the most representative and informative training data.

Further experiments of various training ranges are extensively explored. The comparison shown in Figure 9 indicates that the sand thickness (in orange color), inferred from a deep-learning model training with narrower range of traces, are thinner than the one (in dark blue) training with broader range from the synthetic wedge model. To select the most appropriate number of training set, predicted outputs from each experiment need to be quantitatively validated with the well log information individually. Then, the range is eventually determined and selected purely based on the validation, which provides the inference of sand thickness honoring well logs the best.

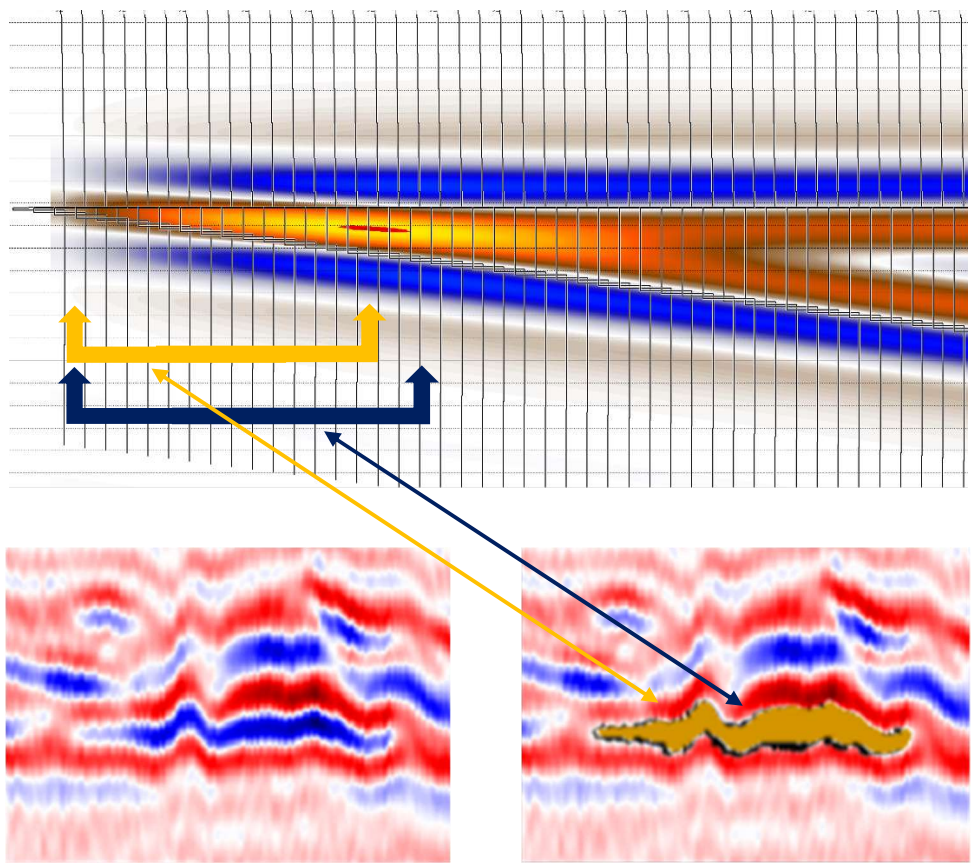

*Figure 9: Comparison of sand thicknesses inferred from two deep-learning models training on two ranges of pairs. (Bottom left indicating the original seismic versus bottom right indicating two inferences overlaid on seismic image.)*

## CONCLUSIONS

This article strongly demonstrates a totally new approach by leveraging a deep-learning model to overcome geophysical limitations in delineation of reservoir characterizations from imperfect seismic data. Even with significant interferences from seismic traces, the proposed approach achieves superior interpretability with a strong link from seismic wavelet to an indicator of sand thickness, which introduces a much easier and more accurate approach in identifying reservoir thickness and characterizations. In a Miocene succession of thinly interbedded sandstones and shales in offshore regions, the proposed approach, which completely removes interferences of wavelets caused by the geological changes vertically and laterally, makes lithologic tracking of reservoirs more straightforward without worrying about over- or under-estimation of seismic net pay for thin beds, where the relationship between original seismic and sand thickness is significantly improved. With fewer interpretation artifacts and interference fingerprints, seismic-guided stratigraphic profiling and depositional-facies could be successfully revealed, when deep-learning technique in conjunction with wedge model are applied.

# ACKNOWLEDGEMENTS

The authors would like to thank Anadarko for the permission to publish this work.